# The unsustainable legacy of the *Nuclear Age* [1]

*The enduring legacy of the* Nuclear Age *is incompatible with the terrestrial (and human) environment*

**Angelo Baracca**

Retired Professor of Physics, Department of Physics and Astronomy, University of Florence, Italy: angelo.baracca@gmail.com

**Keywords**: Nuclear Age; Anthropocene; nuclear tests; radioactive contamination; health consequences; nuclear waste; spent nuclear fuel; plutonium; uranium mining.

In the dispute on the beginning of the *Anthropocene* it has been proposed, among many, a precise date, July 16th 1945, when the *Trinity Test* exploded the first atomic bomb in the desert of Alamogordo[2], which inaugurated the *Nuclear Age*. On the other hand, the almost contemporaneous *Ecomodernist Manifesto* proposed that, among other things, "nuclear fission today represents the only present-day zero-carbon technology with the demonstrated ability to meet most, if not all, of the energy demands of a modern economy."[3]

I do not agree with either of these thesis. The Atomic Age has undoubtedly been a tremendous acceleration of the impact of human activities on natural environment, but in my opinion it joined, however it exacerbated, the trend embarked upon since the First Industrial Revolution, when Capitalism adopted radically new (scientific) methods to exploit and "commodise" Nature and its resources. This breakthrough kicked off the development of industrial processes carried out in physical and chemical conditions further and further away from the conditions of the natural environment on Earth surface, so that they introduced products and procedures which are incompatible with such environment, and therefore produce a permanent and irreversible contamination.[4]

---

[1] A slightly different Italian version of this article was previously published: Angelo Baracca, Antropocene-Capitalocene-Nucleaocene: l'eredità dell'*Era Nucleare* è incompatibile con l'ambiente terrestre (e umano), *Effimera*, September 11th 2018, http://effimera.org/antropocene-capitalocene-nucleocene-leredita-dellera-nucleare-incompatibile-lambiente-terrestre-umano-angelo-baracca/.

[2] Zalasiewicz, J., et al., When did the Anthropocene begin? A mid-twentieth century boundary level is stratigraphically optimal, *Quaternary International*, 383, 2015, pp. 196-203: "We propose an appropriate boundary level here to be the time of the world's first nuclear bomb explosion", on July 16th 1945 at Alamogordo, New Mexico.

[3] *An Ecomodernist Manifesto*, April 2015, https://static1.squarespace.com/static/5515d9f9e4b04d5c3198b7bb/t/552d37bbe4b07a7dd69fcdbb/1429026747046/An+Ecomodernist+Manifesto.pdf. Althoug, honestly, it admits: "However, a variety of social, economic, and institutional challenges make deployment of present-day nuclear technologies at scales necessary to achieve significant climate mitigation unlikely." Unfortunately the desired necessary breakthroughs appear, in my opinion, highly unrealistic: "a new generation of nuclear technologies that are safer and cheaper", apart from being out of time, would imply unsustainable costs for building hundreds of power reactors, and even more the extremely long promised nuclear fusion still appears futuristic, although it absorbs big financing. Neither the times nor the costs or these innovations are compatible with the present environmental and economic conditions.

[4] From this point of view I express some reservation even on the term *Capitalocene*. The outset of Capitalism undoubtedly was an epochal breakthrough in the economic exploitation of Nature, however the majority of the processes developed before the Industrial Revolution were not in irretrievable contrast with the Earth

From this point of view, I have little doubt that the *Nuclear Age* has been an extreme breakthrough in the development of processes and products which are absolutely incompatible with the Earth environment, often really extraneous to it.

My argument has a basic physical reason, namely the enormous energy gap between nuclear and chemical processes, which is of the order of a million times. As a consequence, *the Nuclear Age has produced artificial processes and products that Nature on the Earth is, and never will, absolutely be able to get rid of. The Atomic Age has created problems that have no solution, and will all the more worsen as – civil and military – nuclear technology shall be developed*. Not to mention that nuclear weapons imply the ongoing impending threat of destruction of the human society.

I have always contended that the nuclear choice is a dead-end and no-return way. Once embarked on, it necessarily produces artificial products which can in no way be eliminated, are extremely dangerous for health and the environment (besides perpetuating the risks of military proliferation), and can hardly be disposed in safe and permanent way, isolated form the human society for extremely long times, and requiring in any case huge costs.

It goes without saying that nuclear processes play instead a fundamental role in the Universe, as the "fuel" of stars, at whose interior in fact temperatures of million degrees exist. Nuclear processes are not absent on Earth, but they play a marginal role in physical and chemical phenomena. In any case, "natural radioactivity" entails health and environmental dangers, and is carefully monitored.

A clarification seems necessary. From my considerations the medical applications of nuclear physics are excluded. I don't have sufficient scientific competence, however every physician is aware that any use of radioisotopes or ionizing radiations for therapeutic or diagnostic purposes has unavoidable potential harmful health effects, and the resort to them must be made only in concern-driven cases and with a rigorous cost-benefit assessment. In any case, hospital nuclear waste (although unavoidable) are themselves a serious problem, but are excluded from my analysis.

I would add also that I imagine the possible criticisms to my considerations from some colleagues, whose blind faith in the power of Science and in its ability to solve every problem I well know. Apart deeply disagreeing in principle,[5] here I will only add that any artificial nuclear process designed to remediate the present artificial nuclear products or waste is bound to produce more unstable nuclear products, precisely on account of the energies at which it happens, which are artificially produced on Earth. An example of how one is groping around are past projects of launching nuclear waste in the Sun (or in sea bed): it is not enough that we polluted our Planet! Not to mention the tragic consequences of a possible failure, which occurred in other cases. Anticipating a successive example, the reprocessing of spent nuclear fuel is no solution, it implies the use of huge quantities of highly contaminating substances, and its substantial result is the separation of plutonium, increasing the risks of military proliferation, since it is unrealistic the recycling of such huge quantities of plutonium in nuclear fuel, even in the most optimistic scenario of increase of nuclear power.

In this paper I shall discuss in the order, the devastating and enduring effects of nuclear tests, the burdens and costs of the civil nuclear power programs, the consequences and cleanup of nuclear facilities and nuclear accidents, the unmanageable problem of spent nuclear fuel, the accumulation and dangers of plutonium and military materials, and last but not least the front end of the nuclear cycle, uranium mining, which has always implied the savage exploitation of poor population and countries.

---

environment, they probably could have been reabsorbed. In my opinion it would be methodologically useful to distinguish *Anthropocene, Capitalocene,* and something like *"Industrialcene",* as successive and qualitative increasing steps of the human action on Nature, as distinguished from that of the other living species.

[5] I refer for a deeper discussion to my article "Can Science make peace with the environment? Science, power, exploitation", in T. Arabatzis, J. Renn, A. Simeões, *Relocatig the History of Science***,** Springer, 2015, p. 367-383.

## The unsustainable legacy of nuclear tests … a true nuclear war!

The, increasing, radioactive pollution of the Earth surface and atmosphere by human activities has already caused huge damages, which are not easy to assess: illnesses, contamination form nuclear tests or accidents of wide areas, which sometimes have become practically uninhabitable. The well known environmentalist Rosalie Bertell (1920-2012) edtimated twenty years ago *up to 1,300 millions "victims of the Nuclear Age"*![6] "Killed, maimed or diseased by nuclear power since it's inception. The industry's figures massively underestimate the real cost of nuclear power, in an attempt to hide its victims from the world. … [in order] to shield the nuclear industry from compensation claims from the public". His article is really worth reading.

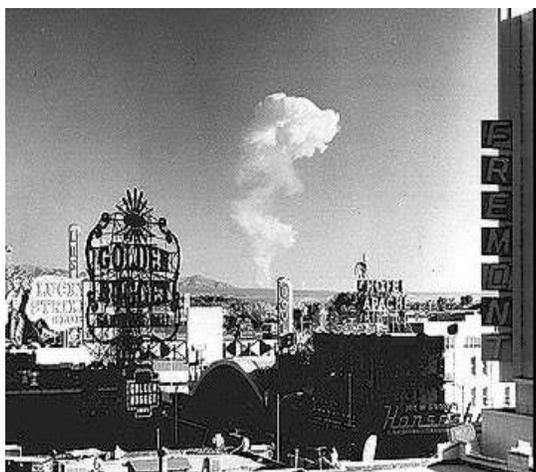

1950s, unscheduled shows in Las Vegas

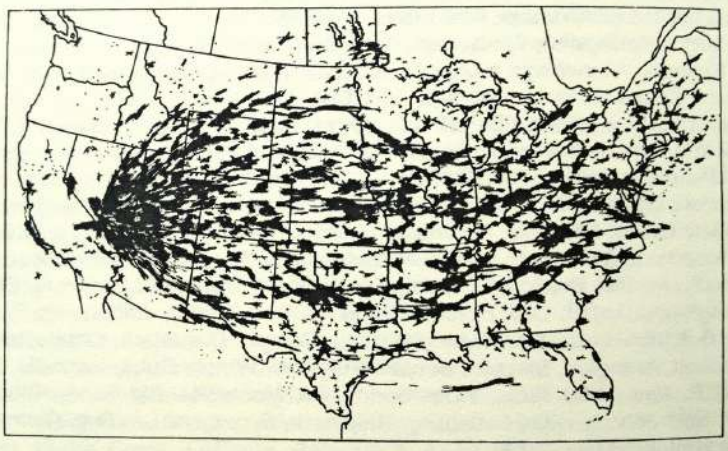

The radioactive pollution produced by nuclear tests, performed from 1945 until 1963 in the atmosphere, reached a maximum value in 1963-64 in the Northern hemisphere, and one year later at the tropics.

In the Nevada desert, in the US, 100 nuclear tests in the atmosphere were performed between 1951 and 1962. Las Vegas is located approximately 50 miles downwind, the mushroom clouds were clearly visible, as well as the glows from Los Angeles.[7] Radioactive pollution spread in the eastern direction on the American territory[8]. "The radionuclide $^{131}$I was one of the main causes of increase thyroid cancer occurrence in the United States, as it was released in large quantities mainly during atmospheric nuclear tests (especially during 1951–1958)".[9] Once the Partial Test Ban Treaty in 1963 formally forbade the nuclear tests in the atmosphere (although some States continued, France until 1995), 921 more subterranean tests were performed. These tests have left an underground residual radioactive contamination of $4.9 \times 10^{18}$ Bq[10] (see details below). Even today radioactive strontium-90 is found in teeth of American children.[11] As well as plutonium in those of British children borne near the nuclear site of Sellafield: needless to say, the British government, not being able to deny, minimizes the consequences.[12]

---

[6] Rosalie Bertell, "Victims of the Nuclear Age", *The Ecologist*, November 1999, p. 408-411, https://ratical.org/radiation/Navictims.html.

[7] Many photographs, not widely known, are readily available in Internet. See for details Harvey Wasserman and Norman Salomon, *Killing our Own: The Disaster of America's Experience with Atomic Radiation*, 1982 (https://www.amazon.com/Killing-our-Own-Experience-Radiation-ebook/dp/B00B10B26Y).

[8] See fallout patterns fir instance in, Nuclear weapons testing, http://www.h-o-m-e.org/weapons/weapons-testing.html. More detailed maps are given in figure 7 of the paper in the following reference.

[9] Remus Prăvălie, Nuclear Weapons Tests and Environmental Consequences: A Global Perspective, *Ambio*, 2014 Oct; 43(6): 729–744, https://www.ncbi.nlm.nih.gov/pmc/articles/PMC4165831/ . In particular see figure 8. Also Nuclear map World National Cancer Institute study estimating thyroid doses of I 131, http://atlantislsc.com/nuclear-map-world/nuclear-map-world-national-cancer-institute-study-estimating-thyroid-doses-of-i-131/

[10] One Becquerel (Bq) is equivalent to one decay per second. All the data are taken from the report *Estimation of*

In the Pacific Ocean 105 nuclear tests in the atmosphere were performed until 1963. Of them 23 in the Bikini Atoll, in the Marshall Islands, between 1946 and 1958. The inhabitants of the Atoll were savagely relocated.[13] These tests produced a tremendous radioactive pollution: tritium $3.4 \times 10^{19}$ Bq, strontium-90 $8.0 \times 10^{16}$ Bq, cesium-137 $1.3 \times 10^{17}$ Bq, plutonium-239 less than $1.0 \times 10^{15}$ Bq. After 60 years the Atoll is still uninhabitable.[14] "The resulting humanitarian needs include recognition, accountability, monitoring, care, compensation and remediation".[15] And to make matters worse, the inhabitants of the Marshall Islands are threatened by the rise of sea level due to climatic change, and could be forced to a new exile.[16]

Soviets were no better in the nuclear test site of Semipalatinsk, in Kazakhstan,[17] where they left a

---

*Global Inventories of Radioactive Waste and Other Radioactive Materials*, IAEA, 2008: https://www-pub.iaea.org/MTCD/Publications/PDF/te_1591_web.pdf.

[11] J. M. Gould, E. J. Sternglass et al., Strontium-90 in deciduous teeth as a factor in early childhood cancer, *Int J Health Serv*, 2000;30(3):515-39, https://www.ncbi.nlm.nih.gov/pubmed/11109179; J. Mangano, "An unexpected rise of Strontium-90 in U.S. deciduous teeth in the 1990s", *The Science of The Total Environment*, Vol. 317 (1-3), December 30, 2003, pp. 37-51 (http://www.radiation.org/); J. Mangano, "Improvements in local infant health after nuclear reactor closing", *Environ. Epid. & Toxic.*, 2 (1-4), 2000; J. Gould, "Explanation of black infant mortality rates", *The Black World Today* (http://www.tbwt.org/home/); D.V. Conn, "US counts one in 12 children disabled", *Washington Post*, 7/6/02; G. Greene, *The Woman Who Knew Too Much: Alice Stewart and the Secret of Radiation*, Univ. Of Michigan Press, 1999; M. L. Wald, Study of Baby Teeth Sees Radiation Effects, *The New York Times*, 13 December 2010, https://www.nytimes.com/2010/12/14/health/14cancer.html; Accumulation of a Radioactive Isotope in Children's Shed Deciduous Teeth Used to Estimate Radiation Exposure from Nuclear Testing and Accidents, Then and Now, *American Dental Association*, March 11, 2016, https://www.ada.org/en/science-research/science-in-the-news/accumulation-of-a-radioactive-isotope-in-childrens-shed-deciduous-teeth.

[12] Plutonium from Sellafield in all children's teeth, 30 November 2003, https://www.theguardian.com/uk/2003/nov/30/greenpolitics.health.

[13] Dramatic photos can be seen in: https://en.wikipedia.org/wiki/Nuclear_testing_at_Bikini_Atoll; https://ww2.kqed.org/wp-content/uploads/sites/10/2017/07/EvacIntoBoat2-1920x1186.jpg; https://www.bikiniatoll.com/Lokiarfamily.jpg.

[14] Bikini Atoll nuclear test: 60 years later and islands still unlivable, https://www.theguardian.com/world/2014/mar/02/bikini-atoll-nuclear-test-60-years.

[15] See a review Tilman A. Ruff, The humanitarian impact and implications of nuclear test explosions in the Pacific region (humbly dedicated to the victims and survivors of nuclear explosions worldwide), *International Review of the Red Cross* (2015), 97 (899), 775–813. *The human cost of nuclear weapons*, file:///home/angelo/Scaricati/irc97_15%20(1).pdf.

[16] Bikini Atoll islanders forced into exile after nuclear tests now find new homes under threat from climate change, 28 ottobre 2015, https://www.independent.co.uk/environment/climate-change/bikini-atoll-islanders-forced-into-exile-after-nuclear-tests-now-find-new-homes-under-threat-from-a6712606.html.

[17] S. Bauer et al., The Legacies of Soviet Nuclear Testing in Kazakhstan. Fallout, Public Health and Societal Issues, *Radioactivity in the Environment*, 19:241-258, January 2013, https://www.researchgate.net/publication/287069767_The_Legacies_of_Soviet_Nuclear_Testing_in_Kazakhstan_Fallout_Public_Health_and_Societal_Issues; B. Grosche et al., Studies of Health Effects from Nuclear Testing near the Semipalatinsk Nuclear Test Site, Kazakhstan, *Cent Asian J Glob Health*, 8 May 2015, https://www.ncbi.nlm.nih.gov/pmc/articles/PMC5661192/; A. Genova, This Is What Nuclear Weapons Leave in Their Wake, *National Geographic*, 13 October 2017, https://www.nationalgeographic.com/photography/proof/2017/10/nuclear-ghosts-kazakhstan/; R. Kobil, Soviet-era nuclear testing is still making people sick in Kazakhstan, *PRI's The World*, 13 March 2017, https://www.pri.org/stories/2017-03-13/soviet-era-nuclear-testing-still-making-people-sick-kazakhstan.

radioactive pollution of 3.5x10$^{15}$ Bq from strontium-90, 6.6x10$^{15}$ Bq from cesium-137, and plutonium-239 less than 1.0x10$^{14}$ Bq.

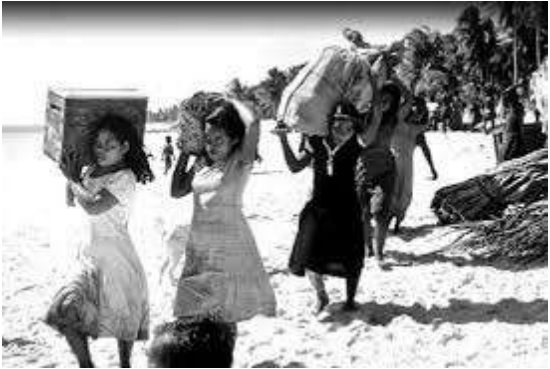
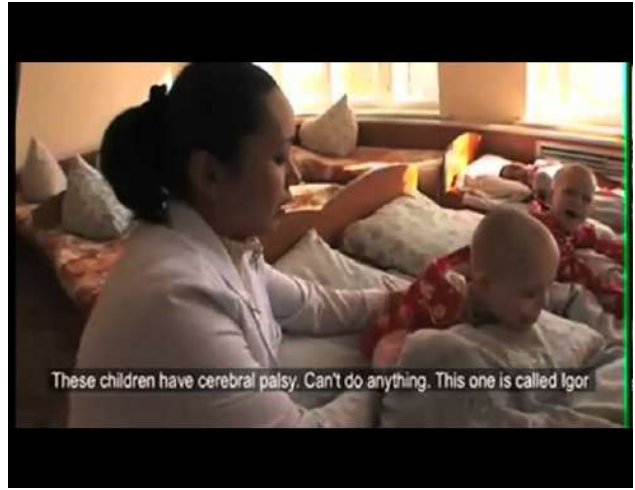

Left: deportation of the people of Marshall Islands

Right: Kazakhstan, health effects of nuclear tests.

From 1960 to 1996, France carried out 210 nuclear tests, 17 in the Algerian Sahara and 193 in French Polynesia in the South Pacific. For decades, France argued that the controlled explosions were clean. By contrast, these tests "were far more toxic than has been previously acknowledged and hit a vast swath of Polynesia with radioactive fallout, according to newly declassified ministry of defense documents which have angered veterans and civilians' groups".[18]

Last but not least, from 1946 through 1993, thirteen nuclear capable countries used the ocean dumping as a method to dispose of nuclear/radioactive waste. "The waste materials included both liquids and solids housed in various containers, as well as reactor vessels, with and without spent or damaged nuclear fuel".[19] The IAEA however claims that "no high-level radioactive waste (HLW) has been disposed of into the sea", but "variable amounts of packaged low-level radioactive waste (LLW) have been dumped at more than 50 sites in the northern part of the Atlantic and Pacific Oceans".[20] Since 1993, ocean disposal has been banned by international treaties (London Convention, 1972; Basel Convention, MARPOL 73/78). The radioactive discharges in the seas from 1946 to 1993 are evaluated in 86x10$^{15}$ Bq.[21]

A letter sent by the recognized specialist Dr. Ernest Sternglass to the Secretary of Energy of the US, Steven Chu, on December 3, 2003, caused a deep shock (unfortunately quickly forgotten). In it Sternglass firmly denounced "a little-known tragic mistake that was made by the medical community and physicists like myself during the early years of the Cold War that has been playing a major role in the enormous rise of the incidence chronic diseases such as cancer and diabetes, and thus the cost of healthcare in our nation. The mistake was to assume that the radiation exposure to the public due to the small amount of fallout from distant nuclear weapons tests or the operation of nuclear reactors would

---

[18] French nuclear tests 'showered vast area of Polynesia with radioactivity', *The Guardian*, 3 July 2013, https://www.theguardian.com/world/2013/jul/03/french-nuclear-tests-polynesia-declassified.

[19] Ocean disposal of radioactive waste, https://en.wikipedia.org/wiki/Ocean_disposal_of_radioactive_waste.

[20] Dominique P. Calmet, Ocean disposal of radioactive waste: Status report, *IAEA Bulletin*, 4 1989, , https://www.iaea.org/sites/default/files/31404684750.pdf. Inventory of radioactive material entering the marine environment: Sea disposal of radioactive waste, *IAEA*, March 1991, p. 47-50, https://www-pub.iaea.org/MTCD/Publications/PDF/te_588_web.pdf.

[21] Ref. 10.

have no significant adverse effect on human health."[22]

This is not the place for going further into the controversial and specialized problem of the health effects of low radiation doses – however crucial it is – but it is worth mentioning that the currently adopted radiation risk model is challenged from a scientific point of view even by the independent experts of the *European Committee on Radiation Risk*[23] (ECRR). Those interested in looking in deep may see the 2010 Report.[24]

*Some remarks regarding underground nuclear test*

Since the Limited Test Ban Treaty of 1963, most explosive nuclear tests have been conducted underground. Nuclear establishments around the world have tried to convince their citizens and others that these tests posed no risks to their health and the environment. This view is questioned. The effects of an underground nuclear test may vary according to factors including the depth and yield of the explosion, as well as the nature of the surrounding rock. When a nuclear bomb explodes underground, the rock surrounding the device is vaporized. Rock lying further from the bomb is melted as temperatures rise by several million degrees. In many cases, the ground above collapses into the molten cavity, allowing radiation to spread into the atmosphere and surrounding environment. Some effects may range from triggering of landslides, tsunamis and earthquakes, physical damage to the reef, venting of gaseous and volatile fission products, medium and long-term leakage of fission products to the biosphere. "For instance, it is estimated that, out of the total number of approx. 800 underground tests performed in the Nevada Test Site, considerable quantities of radionuclide $^{131}$I were released into the atmosphere through venting in at least the 32 known cases of underground tests".[25]

## The enduring, unsustainable burdens and costs of the civil nuclear power programs

Starting from mid-1950s the so-called "civil" nuclear programs for electric energy production were developed. The early exponential increase of these programs began to slow in the 1990s, due to the multiplication of disastrous nuclear accidents (1979 Three Mile Island, 1986 Chernobyl, 2011 no less than four accidents in Fukushima, the main ones). Following these accidents, more severe safety standards were set, and as a result costs enormously grew, as well as construction times: actually, in the last two decades electronuclear production and the construction of new power plants have reached a plateau, and show signs of contraction.

---

[22] Letter from Dr. Ernest Sternglass to Energy Secretary Steven Chu: On health dangers from ingested and inhaled radiation, 3 December 2003, https://healfukushima.org/2014/12/03/letter-from-dr-ernest-sternglass-to-energy-secretary-steven-chu-on-health-dangers-from-ingested-and-inhaled-radiation/.

[23] http://www.euradcom.org/; Introduction to the new website of the European Committee on Radiation Risk, http://euradcom.eu/. The deepness and the relevance of the discrepancies can be appreciated from this excerpt of an ECRR document on the health consequences of the … "It has also been conceded by the editor of the ICRP risk model, Dr Jack Valentin, in a discussion with Chris Busby in Stockholm, Sweden in April 2009. Valentin specifically stated in a videoed interview (available on www.llrc.org and vimeo.com) that the ICRP model could not be used to advise politicians of the health consequences of a nuclear release like the one from Fukushima. Valentin agreed that for certain internal exposures the risk model was insecure by 2 orders of magnitude. The CERRIE committee [see below] stated that the range of insecurity was between 10 and members of the committee put the error at nearer to 1000, a factor which would be necessary to explain the nuclear site child leukemia clusters."

[24] 2010 Recommendations of the ECRR The Health Effects of Exposure to Low Doses of Ionizing Radiation, Edited by Chris Busby with Rosalie Bertell, Inge Schmitz-Feuerhake, Molly Scott Cato and Alexey Yablokov. There is also a *Committee Examining Radiation Risks of Internal Emitters* (CERRIE) which calls for precautionary approach to internal radiation (https://nuclearhistory.wordpress.com/2011/04/25/committee-examining-radiation-risks-of-internal-emitters-cerrie-2001-2004/). Those interested to look more closely at the problem can see the ICRP "2005 Recommendations of the International Commission on Radiological Protection Response from the Low Level Radiation Campaign", http://www.icrp.org/consultation_viewitem.asp?guid=%7BD164F947-85C2-4CD3-AC3E-349AFC52993A%7D.

[25] Remus Prăvălie, Nuclear Weapons Tests and Environmental Consequences: A Global Perspective, cited.

I will not enter into these aspects.[26] Rather, what I want to stress is that *during these 60 years the main concern has been to build new nuclear plants, while huge quantities of the so-called radioactive "waste" piled up* (I don't like the term "waste", since spent fuel contains plutonium and actinides, which have a great military importance, I will return on this). Building was the business. No country in the world has implemented a final storage site for nuclear waste. Spent nuclear fuel and other contaminated material – deadly byproducts of electricity generation – remain stockpiled in temporary locations, sometimes alongside the reactors where they were used.[27] Almost all countries have ongoing projects, generally contested by local population. Some experiences proved to be badly mistaken, if not disastrous. In Germany a deep geological repository for radioactive waste was carried out at Asse, in Lower Saxony, in a former salt mine, but water leakages were subsequently found.[28] More than a hundred thousand barrels of radioactive waste are to be removed, for this disastrous choice the Federal Office is spending €140 million a year for the Asse clean-up.

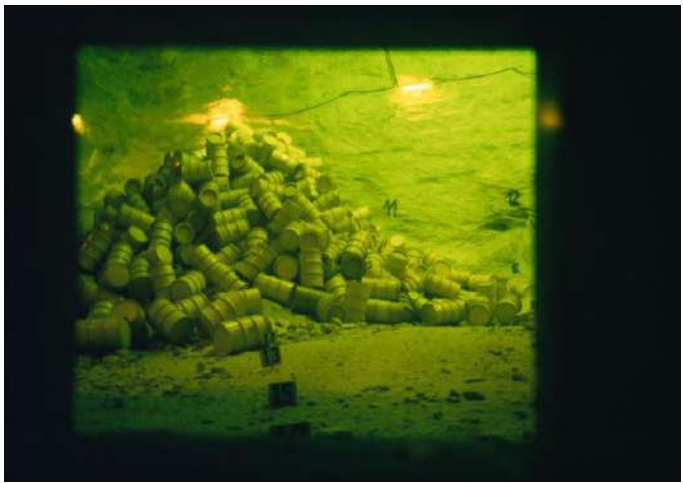
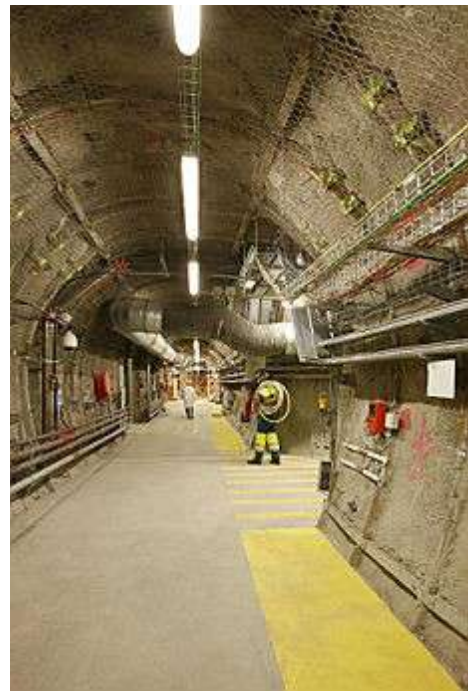

Left: geological dump of Asse, Germany, picture of the 1970s.
Right: works in progress for the final repository of Bure, France (see below).

The frequently used term "final" storage needs a comment. There are various categories of radioactive waste, classified depending on their activity, long-lived waste remain dangerous for centuries or more, so that the term "final" is relative. Radioactive waste must be disposed for a practically indefinite time. The challenge is not only to build a massive radioactive dump but to guard it from human intervention for an impossible amount of time, and against possible future upheavals which could delete the signs and the memory of dumps of dangerous waste.[29]

---

[26] A comprehensive annual up-date is presented by M. Schneider et al., *The World Nuclear Industry Status Report 2017*, Fig. 11, p. 37, https://www.worldnuclearreport.org/-2017-.html.
[27] A recent book by A. Blowers is titled: *The Legacy of Nuclear Power*, Routlege, 2016.
[28] Full history http://www.bfs.de/Asse/EN/topics/what-is/history/history.html.
[29] See for instance A. Marshall, Communicating with future generations about our nuclear waste legacy, *Futures Research Quarterly*, Summer 2007, p. 65-75, https://www.researchgate.net/profile/Alan_Marshall6/publication/306937562_Communicating_With_Future_Ge

Until now, the nuclear waste is commonly stored in repositories considered as "temporary", precisely because a "final" settlement should be provided, at least for a very long time. Thus, enormous quantities of radioactive waste accumulated and continues to grow. The amount of high activity waste in the world, clearly the most dangerous, is estimated around 830,000 cubic meters (those of low or intermediate activity are clearly much more). The title of a recent paper, "Mountains of nuclear waste just keep growing",[30] says it all.

Only in 2011 the EU adopted a rule obliging each country that has produced nuclear waste to have policies on how to manage their waste, all member states were due to report about their national programs for the first time. As of 2013, the United Kingdom has accumulated 154,550 cubic meters of intermediate- and high-level radioactive waste, France 132,200, Germany 25,534.[31]

In this respect a remark is appropriate. Advanced reactor developers pretend that extraordinary waste management benefits can be reaped through adoption of these technologies, and are receiving substantial funding. This thesis is disproved in a recent paper[32] which "describes why molten salt reactors and sodium-cooled fast reactors – due to the unusual chemical compositions of their fuels – will actually exacerbate spent fuel storage and disposal issues. Before these reactors are licensed, policymakers must determine the implications of metal- and salt-based fuels vis a vis the Nuclear Waste Policy Act and the Continued Storage Rule."

*Clean nuclear power does not exist*.

Several countries have ongoing, decades-long, costly projects for the burial of radioactive waste. Finland seems to have the more advanced, and costliest, project, a tunneling project set to cost up to 3.5 billion euros ($5.3 billion) to build and operate until the 2120, when the vaults will be sealed for good.[33] France is committed in a (hotly contested by populations and environmentalists) project near Bure, in northeastern France. An €25bn deep geological storage facility, half a km under the ground, for high and medium-level radioactive waste, whose construction began in 2000 and, if it wins final approval from the French government, should from 2025 be the last resting place. "When the work here is finally finished, no one must ever take this journey again or, at least, not for 100,000 years."[34] France "produces enough toxic radioactive waste every year to fill 120 double-decker buses (about 13,000 cubic meters worth, or 2kg a year for every French person). The challenge at Bure is not only to build a massive dump for radioactive trash but also to guard it from human intervention for an impossible amount of time – more than 4,000 human generations."

## Decommissioning hundreds of shut nuclear plants, another underplayed problem

Besides the accumulation of radioactive waste, even the *decommissioning* of shut down nuclear power plants is going on slowly,[35] while an increasing number of aged plants shall be shut down in the

---

nerations_About_Our_Nuclear_Waste_Legacy/links/57c048ea08ae2f5eb331ed90/Communicating-With-Future-Generations-About-Our-Nuclear-Waste-Legacy.pdf.

[30] Paul Brown, "Mountains of nuclear waste just keep growing", 7 March 2018, https://www.truthdig.com/articles/nuclear-waste-mountains-just-keep-growing/.

[31] European Commission, https://ec.europa.eu/energy/sites/ener/files/documents/staff_working_document_progress_of_implementation_of_council_directive_201170euratom_swd2017_161_final.pdf.

[32] Lindsay Krall & Allison Macfarlane, Burning waste or playing with fire? Waste management considerations for non-traditional reactors, *Bulletin of the Atomic Scientists*, Vol. 74, p. 326-334, 2018.

[33] Finland to bury nuclear waste for 100,000 years in world's costliest tomb, 8 June 2016, http://www.abc.net.au/news/2016-06-08/finns-to-bury-nuclear-waste-in-worlds-costliest-tomb/7488588.

[34] M. Stothard, Nuclear waste: keep out for 100,000 years, *Financial Times*, 14 July 2016, https://www.ft.com/content/db87c16c-4947-11e6-b387-64ab0a67014c.

[35] "Decommissioning nuclear reactors is a long-term and costly process", U.S. Energy Information Administration, 17 November 2017, https://www.eia.gov/todayinenergy/detail.php?id=33792. "To fully decommission a power plant, the

next decades (234 of the 403 world operating nuclear reactors are over 30 years old)[36]. Even this problem has been considered for a long time as a relatively simple and economic task. The MIT nuclear engineering David Rose wrote in the November 1985 issue of *The Bulletin of the Atomic Scientists*: "The most reliable estimate of the cost of decommissioning [a nuclear power plant] is 10-15 percent of the construction cost, contrary to some highly inflated estimates … Modern serious studies of the disposal problem indicate that satisfactory isolation is technologically feasible, even for the long term." However things went very differently, with increasing costs and skyrocketing costs.[37]

National cases are eloquent. According to a recent study, "the clean-up of French reactors will take longer, be more challenging and cost much more than French nuclear operator EDF anticipates."[38] The same happens for the clean-up of the British nuclear program. Since 2013 it was calculated that the bill for cleaning up the huge Sellafield nuclear plant in Cumbria will rise even higher than its current estimated level of £70 bn[39], three years later this estimate was more than doubled,[40] one year ago the latest official estimates put the cost of UK nuclear decommissioning at £164 bn over the next 120 years (that dwarfs the estimated £60 bn cost of decommissioning North Sea oil and gasfields).[41] What will happen in the next 120 years?

Beyond all this, the decommissioning of hundreds of nuclear power plants will add enormous quantities of radioactive waste.

## Cleanup of major power nuclear disasters

The severity of a nuclear accident is absolutely incomparable with any other kind of accidents, because of the unique characteristic of nuclear processes, as I have made clear from the outset. Of

---

facility must be deconstructed and the site returned to greenfield status (meaning the site is safe for reuse for purposes such as housing, farming, or industrial use). Nuclear reactor operators must safely dispose of any onsite nuclear waste and remove or contain any radioactive material, including nuclear fuel as well as irradiated equipment and buildings."

[36] See the already cited survey: M. Schneider et al., *The World Nuclear Industry Status Report 2017*, Fig. 11, p. 37, https://www.worldnuclearreport.org/-2017-.html.

[37] See e.g., L. Song, "Decommissioning a Nuclear Plant Can Cost $1 Billion and Take Decades", *Reuters*, 13 June 2011, https://www.reuters.com/article/idUS178883596820110613. D. Drollette, "The rising cost of decommissioning a nuclear power plant", *The Bulletin of the Atomic Scientists*, 28 April 2014, https://thebulletin.org/2014/04/the-rising-cost-of-decommissioning-a-nuclear-power-plant/. Just an example reported: "The Yankee Nuclear Power Station in Rowe, Massachusetts, took 15 years to decommission—or five times longer than was needed to build it. And decommissioning the plant—constructed early in the 1960s for $39 million—cost $608 million. The plant's spent fuel rods are still stored in a facility on-site, because there is no permanent disposal repository to put them in. To monitor them and make sure the material does not fall into the hands of terrorists or spill into the nearby river costs $8 million per year. That cost will continue for an unknown number of years."
"The global state of nuclear decommissioning: costs rising, funds shrinking, and industry looks to escape liability by decades of delay", *Beyond Nuclear*, 17 April 2016, http://www.beyondnuclear.org/nuclear-decommissioning-costs/2016/4/27/the-global-state-of-nuclear-decommissioning-costs-rising-fun.html.

[38] P. Dorfman, "How much will it really cost to decommission the aging French nuclear fleet?", *EnergyPost*, 15 March 2017, http://energypost.eu/how-much-will-it-really-cost-to-decommission-the-aging-french-nuclear-fleet/, reprinted from *Nuclear Monitor #839*.

[39] "Sellafield executives to face MPs as nuclear clean-up bill rises over £70bn", *The Guardian*, 1 Secember 2013, https://www.theguardian.com/environment/2013/dec/01/sellafield-nuclear-clean-up-cost-rises.

[40] "UK's nuclear clean-up cost estimate dips to $154 billion", 15 July 2016, http://www.world-nuclear-news.org/WR-UK-nuclear-clean-up-cost-estimate-dips-to-154-billion-15071602.html.

[41] A. Ward and G. Plimmer, "UK set to end outsourcing of nuclear clean-up", 15 October 2017, https://www.ft.com/content/b83c5ada-b014-11e7-beba-5521c713abf4. Moreover, moving from outsourcing to in house after contracts collapsed.

course scary "conventional" accidents with long term health and environmental effects occurred (I just mention the enduring consequences of the disaster in Seveso, Italy, in 1976, or in Bhopal, India, in 1984), but major nuclear accidents involving the meltdown of reactor cores renders entire regions practically uninhabitable at least for decades. And the health consequences extend to future generation.

Without going into details,[42] major nuclear accidents (several of them hidden for years, or decades) are worth considering since they eloquently describe the downplaying of both the health and environmental consequences, and the times and costs of cleanup.

I shall consider only the main accidents in civil nuclear power plants, but scary accidents occurred in military plants, I only recall two:

► In 1957 one of the worst nuclear disasters was caused by an explosion in the Soviet plutonium-reprocessing plant in the atomic weapons industry at Mayak, Ural region, which emitted a massive radioactive cloud. The disaster was denounced two years later by the dissident Soviet scientist Zhores Medvedev,[43] but the Soviet government refused until 1989 to acknowledge that the event had occurred, even though about 9,000 square miles (23,000 square km) of land were contaminated, more than 10,000 people were evacuated, and probably hundreds died from the effects of radioactivity. Let me recall that in 2017 Mayak was identified as the likely source of a cloud of a radioactive isotope, ruthenium 106, that was detected over Europe.

► In 1975 a huge accident occurred in the plutonium production plant of Windscale (now Sellafield, Cumbria, see above), in the UK, with a release into atmosphere of radioactive material that spread across the UK and Europe.[44]

*Three Mile Island, 1979* – The first, totally unexpected, severe power nuclear accident occurred 39 years ago (March 28, 1979) at Three Mile Island, Pennsylvania, as Unit 2's reactor partially melted down. "Despite the fact that the Unit 2 reactor coolant system is drained and its radioactive waste was shipped away long ago, the process of decommissioning Unit 2 has not yet begun. But when it does, it will come with an eye-popping price tag: a total of $1.266 billion from 2018 to 2053, according to an analysis released March 26 by the Nuclear Regulatory Commission."[45] For 30 years there were no new plant applications in the US.

The official narrative is that the health effects of the accident were negligible. But "unorthodox" qualified specialists criticize the official thesis. In 1992 the already mentioned Ernest Sternglass asserted: "actually, Three Mile Island caused hundreds of thousands, in fact millions of people in the U.S., to be exposed to the fallout that drifted all across the northern United States and which, in the following year, continued with releases during the venting process when they had to enter the contaminated building. And in the process many thousands of children died prematurely as I documented in the last part of *Secret Fallout*."[46] In 1998 Rosalie Bertell officially denounced in the strongest possible terms the "Ongoing Cover-Up of the Three Mile Island Accident".[47] And in 2017

---

[42] A wide, although partial, list of nuclear power accidents can be found for instance in *Wikipedia*: List of nuclear power accidents by country, https://en.wikipedia.org/wiki/List_of_nuclear_power_accidents_by_country.

[43] Zhores Medvedev, *Nuclear Disaster in the Urals*, 1st Vintage Books ed, New York, 1980.

[44] R. Morelle, Windscale fallout underestimated, *BBC News*, 6 October 2007, http://news.bbc.co.uk/2/hi/science/nature/7030536.stm.

[45] D. Miller, Unit 2 decommission cost is $1.26 billion; it's been 39 years since TMI accident, but process hasn't officially started, *Press&Journal*, 27 March 2018, http://www.pressandjournal.com/stories/unit-2-decommission-cost-is-126-billion-its-been-39-years-since-tmi-accident-but-process,30549.

[46] Ernest Sternglass, Interview, 11 November 1992, https://ratical.org/radiation/inetSeries/ejs1192.html. Sternglass' book is: *Secret Fallout: Low Level Radiation from Hiroshima to Three Mile Island*, McGraw Hill, 1981.

[47] Statement by Dr. Rosalie Bertell, On the Ongoing Cover-Up of the Three Mile Island Accident, July 10, 1998,

Penn State College of Medicine researchers have shown, for the first time, a possible correlation between the Three Mile Island accident and thyroid cancers in the counties surrounding the plant.[48]

*Chernobyl, 1986* – As a premise, the Chernobyl and the Fukushima accidents were radically different, for the characteristics of the contamination and the landscapes affected, the radiological criteria, the designation of areas to be remediated and the remediation measures adopted.[49]

On the health consequences of the Chernobyl disaster and the present situation the reassuring conclusions of the official international agencies are in blatant contrast with the analysis of independent organizations, like *Greenpeace*. In fact, in 2003 and 2005 the UN International Atomic Energy Agency Chernobyl Forum report predicted 4,000 additional deaths attributable to the accident.[50] In 2006 a report by *Greenpeace*, which involved 52 respected scientists, denounced the IAEA reports as a gross simplification of the real breadth of human suffering: "The new data, based on Belarus national cancer statistics, predicts approximately 270,000 cancers and 93,000 fatal cancer cases caused by Chernobyl. The report also concludes that on the basis of demographic data, during the last 15 years, 60,000 people have additionally died in Russia because of the Chernobyl accident, and estimates of the total death toll for the Ukraine and Belarus could reach another 140,000."[51]

Just two years ago Yury Bandazhevsky – former director of the Medical Institute in Gomel (Belarus), working on sanitary consequences of the Chernobyl disaster – flatly declared "Chernobyl is not finished, it has only just begun".[52]

The overall cost of Chernobyl nuclear disaster could be of hundreds of billions of dollars,[53] taking into account the direct damage, the cost of sealing off the reactor, the creation of the exclusion zone,

---

https://ratical.org/radiation/RbonTMIcu.html.

[48] D. Goldberg et al., Altered molecular profile in thyroid cancers from patients affected by the Three Mile Island nuclear accident, *The Laryngoscope*, Vol. 127, June 1, 2017,
https://onlinelibrary.wiley.com/doi/abs/10.1002/lary.26687.

[49] S. Nakayama et al., A comparison of remediation after the Chernobyl and Fukushima daiichi accidents, 14th International Congress of the International Radiation Protection Association (IRPA) Cape Town, South Africa May 9-13, 2016, https://gnssn.iaea.org/RTWS/general/Shared%20Documents/Environmental%20Assessment/TM-52829%2013-17%20June%202016/Presentations%2016%20June%202016/36-Nakayama(Japan)_Comp%20remediation%20Chern%20Fuku.pdf. B. J. Howard et al., A Comparison of Remediation After The Chernobyl and Fukushima Daiichi Accidents, *Radiation Protection Dosimetry*, Volume 173, Issue 1-3, 1 April 2017, Pages 170–176, https://academic.oup.com/rpd/article-abstract/173/1-3/170/2585097?redirectedFrom=fulltext.

[50] Chernobyl's Legacy: Health, Environmental and Socio-Economic Impacts, The Chernobyl Forum: 2003–2005, https://www.iaea.org/sites/default/files/chernobyl.pdf.

[51] Chernobyl death toll grossly underestimated, *Greenpeace International*, 18 April 2006, https://www.greenpeace.org/archive-international/en/news/features/chernobyl-deaths-180406/.

[52] K. Hjelmgaard, Exiled scientist: 'Chernobyl is not finished, it has only just begun', *USA Today*, 18 April 2016, https://eu.usatoday.com/story/news/world/2016/04/17/nuclear-exile-chernobyl-30th-anniversary/82896510/.
Bandazhevsky created an institute in Belarus, in 1989, specially dedicated to study Chernobyl's impact on people's health, particularly children. In 1999 he was arrested in Belarus and sentenced to eight years in prison for allegedly taking bribes from parents trying to get their children admitted to his Gomel State Medical Institute. He denied the charges. The National Academy of Sciences and Amnesty International say he was detained for his outspoken criticism of Belarus' public health policies following the nuclear disaster. He was released in 2005 and given French citizenship. He has not returned to Belarus for fear that his family there could be persecuted or arrested by authorities, and now runs a medical and rehabilitation center outside Kiev dedicated to studying and caring for Chernobyl's victims.

[53] K. Amadeo, The Chernobyl Nuclear Power Plant Disaster and Its Economic Impact, *the balance*, 11 April 2018, https://www.thebalance.com/chernobyl-nuclear-power-plant-disaster-economic-impact-3306335.

the resettlement of 330,000 people, health care for those exposed to radiation, seven million people who are still receiving benefits payments in Russia, Ukraine, and Belarus, and so on.

*Fukushima, 2011* – In short, the tsunami triggered the meltdowns in Units 1, 2, and 3, and serious damage of a decontamination pool (see below), *4 major accidents at once*. Subsequent explosions caused by hydrogen buildup (from zirconium cladding of fuel assemblies melting and oxidizing) in Units 1, 3, and 4 then expelled radioactive contamination, most of which fell within the confines of the plant.

After more than seven years from the accident there is no end in sight for recovery.[54] Removing nuclear fuel from the Fukushima Daiichi power plant will take 30 to 40 years, and Japan faces myriad challenges to decommissioning and decontamination. The damaged reactors must still be cooled pumping water inside, although the amount of contaminated water that must be pumped out and treated every day has decreased significantly. Moreover, every day, as much as much as 150 tons of groundwater percolates into the reactors through cracks in their foundations, becoming contaminated with radioactive isotopes in the process. The ever-increasing volumes of contaminated water is stored in hundreds of temporary tanks, which contain 1 million of the 1.1 million-ton total capacity, including 850,000 tons of so-called tritiated water (the highly contaminated coolant water that directly contacted the fuel is sent to filtering systems and purified into a mix of water and tritium, a radioactive form of hydrogen that is difficult to separate from water). Obviously it is not realistic to permanently keep this water in tanks, a new tank is being added weekly.[55] Roughly 1,000 tons of polluted water were dumped into the sea after a typhoon hit the facility.[56]

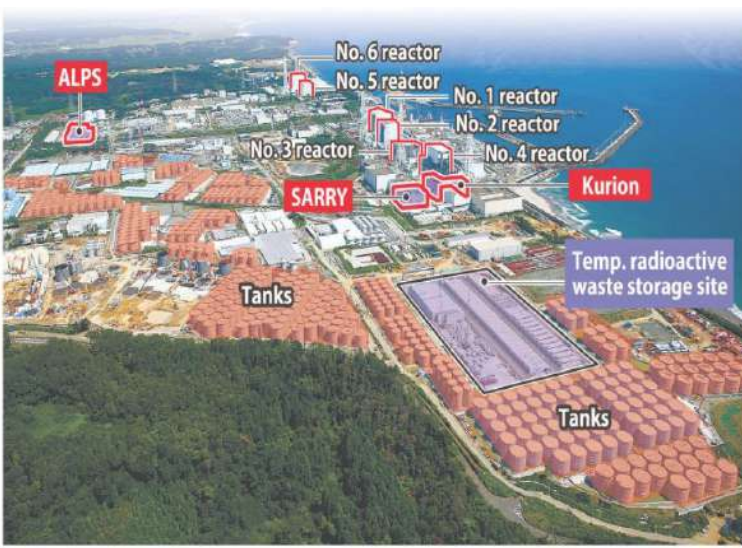

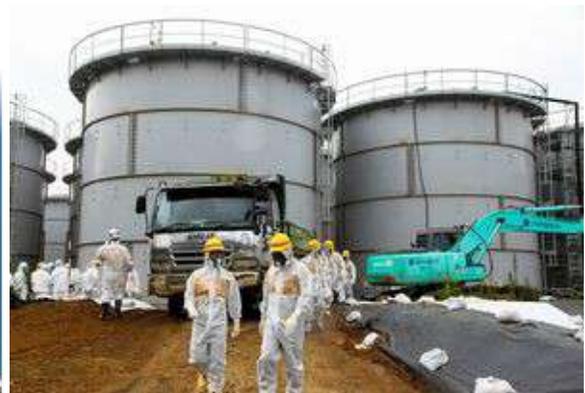

Fukushima, tanks of contaminated water

---

[54] S, Muramatsu & K. Hanawa, Seven years on, no end in sight for Fukushima's long recovery, *Nikkei Asian Review*, 11 March 2018, https://asia.nikkei.com/Economy/Seven-years-on-no-end-in-sight-for-Fukushima-s-long-recovery. Murai, Fukushima No. 1 cleanup continues but radioactive water, and rumors, also prove toxic, *The Japantiomes*, 9 March 2018, https://www.japantimes.co.jp/news/2018/03/09/national/fukushima-no-1-cleanup-continues-radioactive-water-rumors-also-prove-toxic/#.W6_ue87fVpo.

[55] J. Sturmer, Japan undecided on what to do with 1 million tonnes of radioactive water at Fukushima plant, *ABC News*, 2 March 2018, http://www.abc.net.au/news/2018-03-02/fukushimas-radioactive-water-still-a-dilemma-for-japanese-gov/9504072.

[56] Fukushima nuclear plant dumps 1,000 tons of polluted water into sea, *The Telegraph*, 17 September 2013, https://www.telegraph.co.uk/news/worldnews/asia/japan/10314444/Fukushima-nuclear-plant-dumps-1000-tons-of-polluted-water-into-sea.html.

Remote-controlled robots have provided a limited view of melted fuel debris inside the reactors, and finally found the melted uranium fuel inside Unit 3.[57] Still, the exact location of the melted fuel is largely unknown and robots that can withstand the high radiation for prolonged work there are still being developed. Among the highest risks at the plant are 1,573 fuel rod units, each consisting of dozens of fuel rods, which are cooled with water in storage pools (see below) that are not enclosed within the reactor buildings. Recently, uranium and other radioactive materials, such as cesium and technetium, have been found in tiny particles released from the damaged Fukushima Daiichi nuclear reactors.[58]

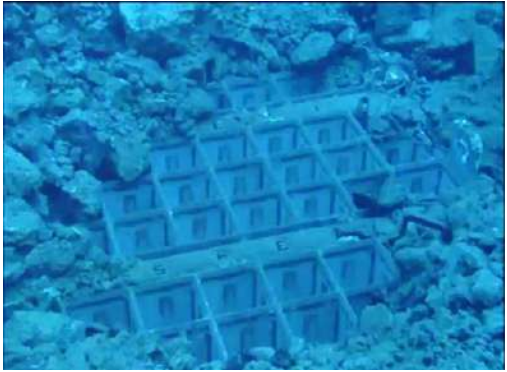
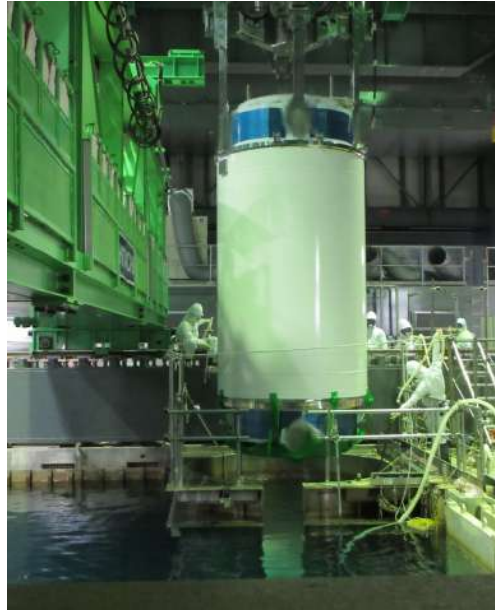

Left: damaged spent fuel rod pool Unit 4.
I recommend to to watch the video:
https://www.youtube.com/watch?v=QVqfPCsl2AA
Right: removal of spent fuel rods from reactor 4 pool into casks (see below).

In 2016 the direct costs of the Fukushima disaster were estimated in about $15 billion in clean-up over the next 20 years and over $ 60 billion in refugee compensation.[59] Replacing Japan's 300 billion kWhs from nuclear each year with fossil fuels has cost japan over $ 200 billion, mostly from fuel costs for natural gas, fuel oil and coal, as renewables have failed to expand in Japan. This cost will at least double, and that only if the nuclear fleet could be mostly restarted by 2020. The reconstruction and recovery costs associated with just the earthquake and the tsunami will top $250 billion. Almost three years ago, Japan's government nearly doubled its projections for costs related to the Fukushima nuclear disaster to $188 billion.[60]

The precise value of the abandoned cities, towns, agricultural lands, businesses, homes and property located within the roughly 310 sq miles (800 sq km) of the exclusion zones has not been established. Estimates, in 2012, of the total economic loss range from $250 to $500 billion.[61]

---

[57] M. Fackler, Six Years After Fukushima, Robots Finally Find Reactors' Melted Uranium Fuel, *The New York Times*, 19 Novembner 2017, https://www.nytimes.com/2017/11/19/science/japan-fukushima-nuclear-meltdown-fuel.html.
[58] New evidence of nuclear fuel releases found at Fukushima, *Science Daily*, 28 February 2018, http://www.abc.net.au/news/2018-03-02/fukushimas-radioactive-water-still-a-dilemma-for-japanese-gov/950407.
[59] James Conca, After five years, what is the cost of Fukushima?, *Forbes*, March 10 2016, https://www.forbes.com/sites/jamesconca/2016/03/10/after-five-years-what-is-the-cost-of-fukushima/#1b32931522ed.
[60] Yuka Obayashi and Kentaro Hamada, Japan nearly doubles Fukushima disaster-related cost to $188 billion, , December 9, 2016, *Reuters*, https://www.reuters.com/article/us-tepco-fukushima-costs/japan-nearly-doubles-fukushima-disaster-related-cost-to-188-billion-idUSKBN13Y047.
[61] NewsonJapan.com, Fukushima cleanup could cos up to $250 billions, *NewsOnJapan.com,* 6 Nov. 2012; Amie Gundersen & Helen Caldicott, The ongoing damage and danger at Fukushima, *Fairewinds Energy*

*When the next nuclear accident?* – This question is not a voice of doom. It is certain that every machine, even the most perfect one, is subject sooner or later to breakdown and malfunctions. The more a machine is sophisticated, the more it is subject to accidents, and the more it is difficult to predict every possible cause. But above all it is impossible to predict *when* a specific accident will occur. The usual philosophy is to calculate the *probability* for a specific accident to occur. The probabilities for big nuclear accidents have always been predicted, with complex assumptions and criteria, as very small, and this was assumed as an estimate of very low frequencies. Needless to say, the forecasts have always been updated after every accident.

I am not able to evaluate the complex criteria of the calculation of these probabilities, but in the first place I would dispute the interpretation of probabilities, even if they were very small, assumed as a reliance that big nuclear accidents would extremely hardly ever happen. Consider the following example, a macroscopic sample of natural uranium, which is composed for 99,7% by the isotope 237 with a half-life of about 4 billion years, which means that each atom has an extremely small decay probability (half-life is the inverse of the decay probability). Actually, if we put a Geiger counter near the sample, it signals an almost continuous rate of decays. The nuclei which decay now do not wait 4 billion years (although they have an extremely low decay probability), on the other hand other nuclei will not decay during the full life of the universe. It is true that the number of nuclei contained in a macroscopic sample is extremely high (of the order of $10^{23}$), however this is the true meaning of probability, in no way it does indicate *when* an event really occurs. This is crucial in case of events with catastrophic consequences, an even (supposed) very low probability does not authorize to rest assured. As a minimum the danger should take into account both the probability and the severity of the effects. An extremely serious accident should *never* occur!

What's more, every calculation of the probability of an event is model dependent. In the case of very complex events there are many assumptions, and many factors which cannot be taken into account. Actually, after every nuclear accident, showing the role of new factors, the calculation of the probability has changed.

Considering only the major accidents, counting four accidents in Fukushima, 6 severe events occurred since 1979, from which a practical evaluation is inferred of one serious accident every 7 years. I assume that this is a more reliable criterion than probability. Caution must be absolute, and prevail over any assurance from technicians and nuclear industry.

It is absolutely crazy to rely on the confidence in nuclear technology. The world fleet of power nuclear reactors have an average age of 29.3 years. Owing to the present standstill in the construction of new nuclear power plants, the life of the plants into operation is being extended over the original design of 40 years, up to 60. This is a Russian roulette. Every machine growing old is more subject to accident. The question is not *if* but *when*!

## Spent nuclear fuel, a never-ending story

Almost 180,000 tons of spent fuel have accumulated in the world.

Once extracted from the nuclear plant spent fuel must be cooled in the spent fuel pools for at least one year and often much longer, and after this cooling it can be treated basically in two ways. The first, and the more common one is the so-called *dry cask storage* of spent fuel as such. The casks are typically steel cylinders that are either welded or bolted closed. The fuel rods inside are surrounded by inert gas. The steel cylinder provides a leak-tight confinement of the spent fuel. Each cylinder is surrounded by additional steel, concrete, or other material to provide radiation shielding to workers and members of the public. In any case, these casks must be stored in a waste facility which guarantees the absolute external insulation and absence of water leakages for at least centuries. Here lies the problem.

---

*Education.*Web, 6 Nov. 2012.

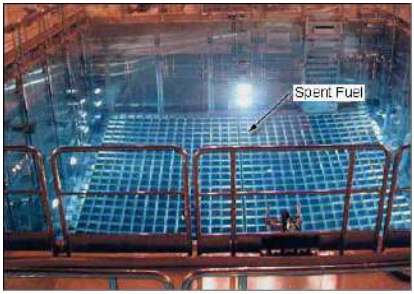
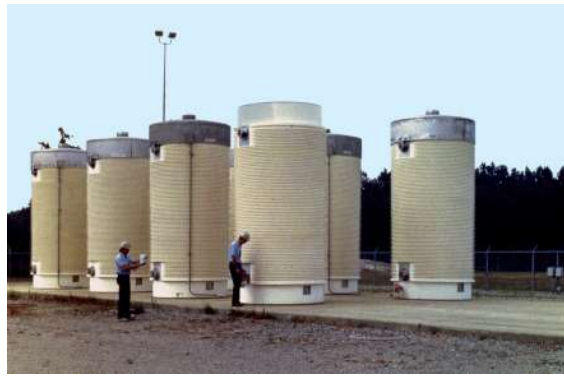
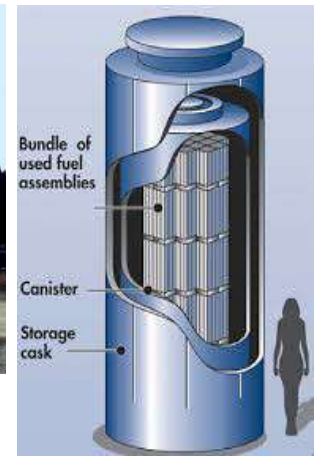

Left: pool for spent fuel.
Center: dry cask storage.
Right: section of a cask.

No country in the world has yet carried out a final repository for spent fuel, although many projects are underway. The United States conducted a big project of a deep geological repository storage facility within Yucca Mountain for spent nuclear fuel and other high-level radioactive waste, located on federal land adjacent to the Nevada Test Site about 80 mi (130 km) northwest of the Las Vegas Valley. This project was designated by the Nuclear Waste Policy Act amendments of 1987, it was approved in 2002, it encountered many difficulties and was widely opposed in Nevada, in particular by the Shoshone peoples (for whom this is sacred land), and it was suspended in 2011 under the Obama administration. This leaves American utilities and the United States government without any designated long-term storage site for the high-level radioactive waste stored on site at various nuclear facilities around the country. At present transuranic waste is currently disposed 2,150 feet (660 m) below the surface at the Waste Isolation Pilot Plant (WIPP) in New Mexico.

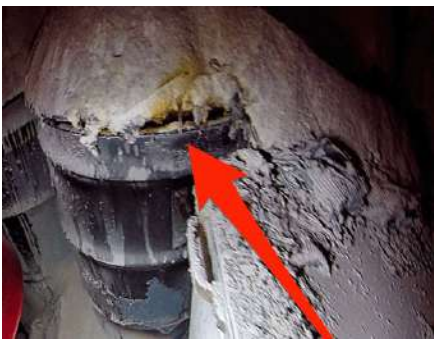
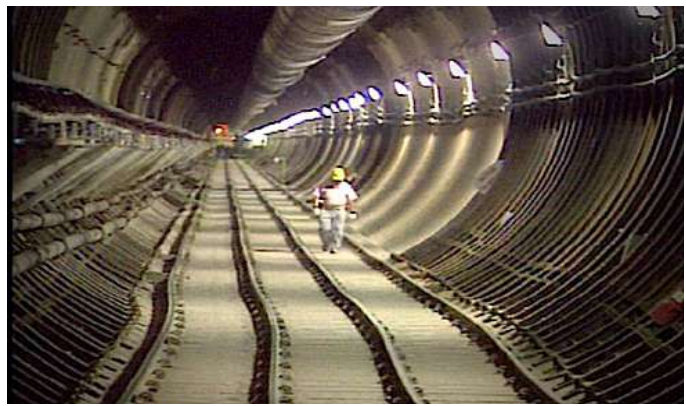

Left: exploded drum at the WIPP.
Right: works at Yucca Mountain

Alternatively to dry storage, spent fuel can be *reprocessed*, resulting in the separation of plutonium generated in the chain reaction. Plutonium is a transuranic nucleus which does non exist in nature,[62] but is the ideal nuclear "explosive". It presents therefore major risks of military proliferation. Actually, plutonium can be mixed with uranium in the so called Mox (mixed oxide) fuel, but on the one hand is seems completely unrealistic that the huge quantities accumulated (see below) could be used in this way by the stagnant nuclear programs, and in any case more plutonium is being produced than is being recycled as reactor fuel. It is moreover worth pointing out that reprocessing involves extremely dangerous and polluting processes, and produces a greater amount of radioactive and toxic waste.

---

[62] Pu-244 was found in Precambrian Age phosphate from southern California. This isotope of plutonium had a radioactive half-life of 80 million years. Scientists have postulated that, because of its long radioactive half-life, this isotope has existed since the creation of Earth about 4.5 billion years ago.

Since the end of the 1970s the United States, under the presidency of Jimmy Carter (who was a nuclear engineer), abandoned reprocessing of spent fuel, precisely for proliferation concerns. Among the western counties, only France is still pursuing reprocessing of spent fuel (even on behalf of other countries).

Needless to say, reprocessing of spent fuel from nuclear reactors is one of the paths for countries which went nuclear (so it was, for instance, for Israel, India, North Korea).

At present around 90,000 metric tons of spent fuel in the United States are stored approximately one half in pools, and one half in casks. For 2050 this quantity is expected to double, while the storage in pools should be progressively eliminated. However at the moment the US seems in the dark, after the stop to the project of Yucca Mountain geological waste facility.

In this connection, it is important to remark that the activity of spent fuel takes extremely long times to decrease[63]. Initially its activity is approximately 1 million times the mining of natural uranium ore. Fission and activation products take 1,000 years to drop below this value, but actinides and actinides daughters take 100,000 years! Will human species, or civilization, survive so long?

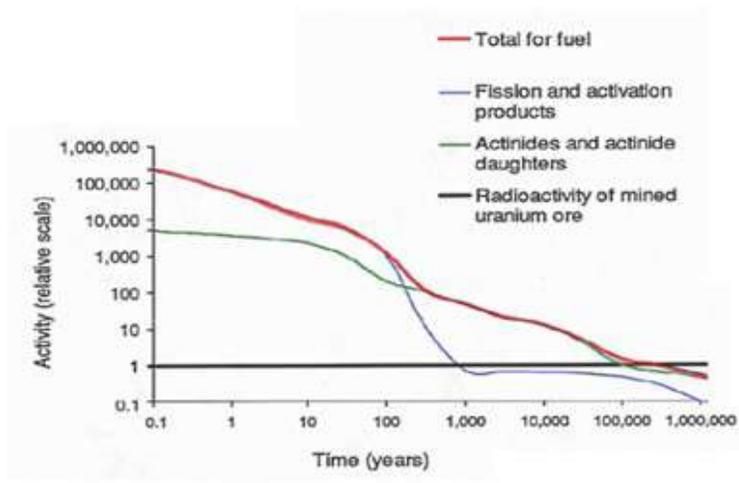

Time decay of the activity of spent fuel, and its byproducts compared with the activity of uranium ore.

In short, *we have generated extremely dangerous artificial processes and products which cannot be eliminated, at most they must be isolated (if possible) from human activities, and presumably they could survive to the disappearance of the human species*. If humankind will survive for some millennium, how pass down to future generations the information on sites which house extremely hazardous materials? In spite of cataclysmic events, or wars?

**Military nuclear materials, plutonium**

Nuclear military activities have produced almost 100-200 million tons of radioactive waste,[64] of which 400,000 tons of depleted uranium, $10^{20}$ Bq of high activity, and $7 \times 10^{17}$ Bq of low activity waste.

The case of the Hanford decommissioned nuclear production complex, in the United States, may serve as an eloquent example. This complex, where the plutonium of *Trinity* and *Fat Boy* was made, covers 586 square miles, nuclear waste is stored at 1200 sites, there are 43 million gallons of high-level waste, 25 million cubic feet of solid radioactive waste, and beneath the site lie 200 square miles of contaminated groundwater.[65] About 100 million gallons are stored in 227 underground tanks, many

---

[63] Nuclear Storage & Disposal, *Nuclear waste concern*, 31 March 2016, http://nuclearwasteconcerns.blogspot.com/2016/03/nuclear-storage-disposal.html

[64] A. Glaser e Zia Mian, "Fissile material stocks and production, 2008", *Bulletin of the Atmic Scientists*, January/February 2009, p. 25-47, https://www3.nd.edu/~dlindley/handouts/Fissile%20Stockpiles.pdf.

[65] Taylor Kate Brown, 25 years on at America's most contaminated nuclear waste site, *BBC News*, 11 June 2014,

larger than state capitol domes and ranging in age from 43 to 73 years. Over 1 million gallons of these contaminants have leaked at the DOE's Hanford site in Washington state, threatening the Columbia River. Undocumented contamination continue to be discovered. The total waste inventory is unknowable but is probably about $1.3 \times 10^{19}$ Bq and about 400,000 tons chemicals. A new discovery, in May 2017, was a 20-foot-by-20-foot hole in the roof of a tunnel built in 1956, and sealed in 1965, where rail cars carrying high concentrations of nuclear waste are buried.[66]

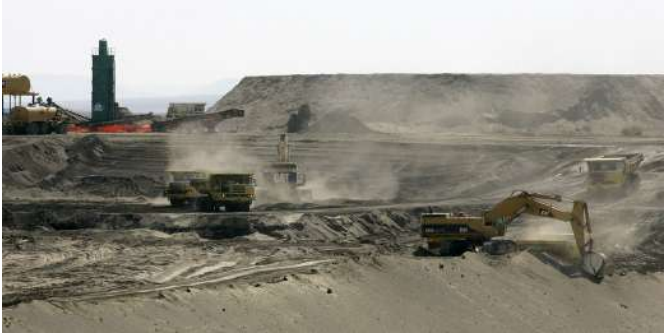
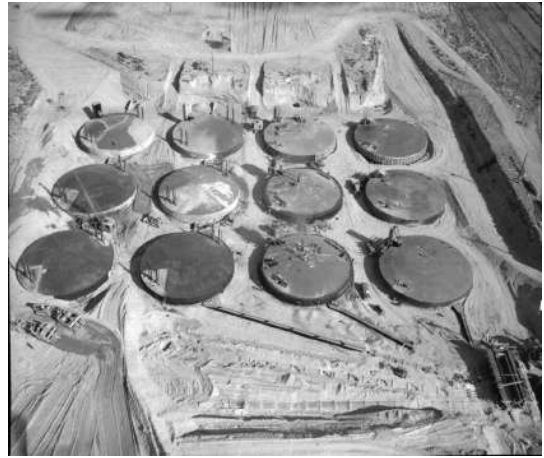

Hanford. Left: 2008, burial of radioactive debris.
Right: partial view of the "tank-farm", tanks containing radioactive waste.

With regard to the costs for Hanford cleanup, the last estimate made public in a lifecycle report put the remaining cost of environmental cleanup of the nuclear reservation at $107.7 billion.[67] However, "With its $1.2 trillion price tag for the modernization of the U.S. nuclear weapons arsenal and production complex, the U.S. Congressional Budget Office has induced 'sticker shock' on Capitol Hill."[68] "The largest of these cleanup costs, at $179.5 billion, is attributed to the stabilization and disposal of high-level radioactive wastes generated from the production of plutonium."

Moreover, "The United States is already paying a stiff price for the harm caused to the workers who made nuclear weapons through the 1980s. To date, 120,599 deceased and sick nuclear weapons workers have been paid $ 15.37 billion in compensation and medical care."[69] The article denounces that the Energy Department is seeking to block the Defense Nuclear Facilities Safety Board's access to critical safety information. "The Energy Department manages the US nuclear weapons complex in an unusual manner."

On the Kola Peninsula in northwest Russia, within the Arctic Circle, lies one of the most bleak naval facilities on the planet, an incredible Soviet submarine graveyard.[70] The rusting submarine hulks reportedly date back to the 1970s. During that time, shipyards struggling to keep up with large military orders didn't have the resources to decommission and dismantle older vessels. Photos of

---

https://www.bbc.com/news/magazine-26658719.
[66] T. Pittman, Sinkhole filled at collapsed Hanford nuke site tunnel, *USA Today*, 12 May 2017, https://eu.usatoday.com/story/news/nation-now/2017/05/12/sinkhole-collapsed-hanford-nuke-site/319098001/.
[67] Annette Cary, DOE gets another pass on estimating Hanford cleanup costs, *Tri-City Herald*, 12 September 2017, https://www.tri-cityherald.com/news/local/hanford/article172985661.html.
[68] Robert Alavarez, The cost of cleaning up our nuke weapons waste is soaring, *Newsweek*, 1 February 2018, https://www.newsweek.com/cost-cleaning-our-nuke-weapons-waste-soaring-767006.
[69] Robert Alvarez, Under siege: Safety in the nuclear weapons complex, *Bulletin of the Atomic Scientists*, August 30 2018, https://thebulletin.org/2018/08/under-siege-safety-in-the-nuclear-weapons-complex/?utm_source=Bulletin%20Newsletter&utm_medium=iContact%20Email&utm_campaign=August31.
[70] Forgotten Soviet Submarine Graveyard on the Kola Peninsula, 2011, https://www.urbanghostsmedia.com/2011/08/forgotten-soviet-submarine-graveyard-kola-peninsula/.

decommissioning of Soviet submarines are really amazing.[71]

The former Soviet submarine graveyard is not an isolated example. Two graveyards of submarines exist in Britain.[72] They rise safety concern.[73] And obviously in the Unites States.[74]

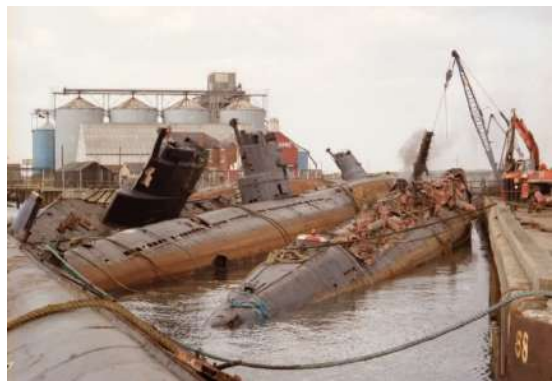 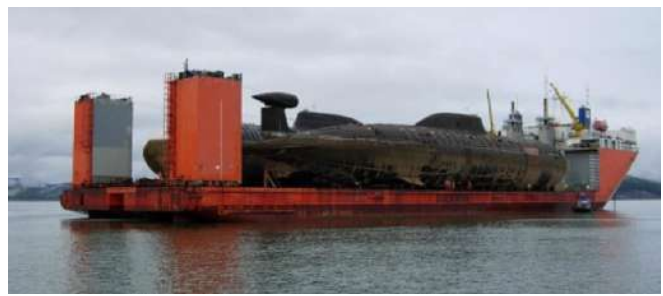

Left: Soviet submarine graveyard.
Right: transport of Russian submarine for dismantling.

The list of world sunken nuclear submarines is impressing, they are ticking rime-bombs hidden into the oceans![75] As well as the list of military nuclear accidents.[76] Dozens of nuclear warheads have been lost at sea by superpowers.[77] Two of the most known disasters are the following. The USS Scorpion nuclear submarine sank on May 22, 1968, southwest of the Azores, with torpedoes, some of which may have carried nuclear warheads, all 99 men on board died.[78] The Soviet submarine sank in the Bay of Biscay in the northeast Atlantic Ocean (where in late World War II British and American aircraft sank nearly seventy German U-boats) on April 12, 1970, propelled by two nuclear reactors, and armed with four torpedoes tipped by nuclear warheads.[79]

---

[71] Here Are Amazing Photos Of Russia Dismantling An Outdated Nuclear Submarine, *Business Insider*, 10 October 2014, https://www.businessinsider.com.au/how-russia-dismantles-its-old-nuclear-subs-2014-10; decommissioning of a Viktor, 18 August 2015, http://www.betasom.it/forum/index.php?showtopic=44611; УНИЧТОЖЕНИЕ ФЛОТА РОССИИ В ЕЛЬЦИНОВСКУЮ ЭПОХУ, *Swalker.ru*, https://swalker.org/voennie/444-kak-my-podnyali-vmf-rf-s-kolen-i-pustil-na.html.

[72] J. Morris, Devonport: Living next to a nuclear submarine graveyard, *BBC News*, 2 October 2014, https://www.bbc.com/news/uk-england-devon-28157707. J. Morris, Laid-up nuclear submarines at Rosyth and Devonport cost £16m, *BBC News*, 3 June 2015, https://www.bbc.com/news/uk-england-devon-32086030. The painfully slow process of dismantling ex-Royal Navy nuclear submarines, *Save the Royal Navy*, 10 January 2018, https://www.savetheroyalnavy.org/the-painfully-slow-process-of-dismantling-ex-royal-navy-nuclear-submarines/.

[73] R. Edward, Questions raised over safety regime at Scotland's nuclear submarine graveyard, *The Herald*, 10 January 2016, https://www.heraldscotland.com/news/environment/14194001.questions-raised-over-safety-regime-at-scotlands-nuclear-submarine-graveyard/; Devonport and nuclear submarines: what are the risks?, *Scientists for Global Responsibility*, Newsletter N. 45, 24 February 2017, http://www.sgr.org.uk/resources/devonport-and-nuclear-submarines-what-are-risks.

[74] J. Talton, Submarines dismantled in Puget Sound are symbols of nation's defense dilemma, *The Seattle Times*, 27 May 2017, https://www.seattletimes.com/business/submarines-dismantled-here-are-symbols-of-nations-defense-dilemma/.

[75] List of sunken nuclear submarines, https://ipfs.io/ipfs/QmXoypizjW3WknFiJnKLwHCnL72vedxjQkDDP1mXWo6uco/wiki/List_of_sunken_nuclear_submarines.html.

[76] List of military nuclear accidents, https://ipfs.io/ipfs/QmXoypizjW3WknFiJnKLwHCnL72vedxjQkDDP1mXWo6uco/wiki/List_of_military_nuclear_accidents.html.

[77] A. Rosenthal, Dozens of Atomic Warheads Lost In Sea by Superpowers, Study Says, *The New York Times*, 7 June 1989, https://www.nytimes.com/1989/06/07/us/dozens-of-atomic-warheads-lost-in-sea-by-superpowers-study-says.html.

[78] The USS Scorpion, Mystery of the Deep, 21 May 1998, http://northwestvets.com/spurs/scorpion.htm

[79] R. Farley, A Dead Russian Submarine Is Sitting on the Bottom of the Ocean (Armed with Nuclear Weapons).

*Depleted uranium* – Depleted uranium is a by-product of the enrichment process, its military use is well known since the Gulf War of 1989[80] in what could be called "low intensity" radiological wars, that caused increases of cancers and other illnesses. Not just among civil populations living in the theaters of war, but also among the unaware troops, like the Italian ones that served in Bosnia, which already suffered almost 350 deaths.[81]

## *Plutonium*

Plutonium deserves some specific remarks. As I have said, this artificial transuranic element is the ideal nuclear "explosive", in 1942 the Fermi "pile" was built with the goal of verifying its feasibility, and successive nuclear reactors were built for military purpose, plutonium production.

Almost 1.300 tons of plutonium have been produced up today[82] (not including that contained in non reprocessed spent fuel, as well as in the *pits* of almost 15,000 still intact nuclear warheads), of which 260 tons *weapon grade*, that is directly suitable for nuclear weapons. The remaining almost 1,000 tons are *commercial grade* plutonium, but it has been shown since long time that nuclear weapons may be made from every kind of plutonium.[83] In general it has been remarked that "The Light Water Reactor (LWR) ... is not nearly so "proliferation resistant" as it has been widely advertised to be. From a proliferation point of view the LWR is generally preferable to other types of power reactors but the differences are more blurred than was previously appreciated."[84]

What to do with such huge amounts of plutonium? Note that plutonium is the most toxic known material, both from the radioactive and the chemical point of view. The only solution for it (apart the limited possibility of fabricating Mox fuel) is to store it in the most possible safe way! Protecting it against thefts for military use, mainly in certain countries. Plutonium-239 has a half life is roughly 24,000 years. *One more extremely dangerous legacy of the Atomic Age which cannot be eliminated, and is practically perennial, on the human scale*.

Japan is worthy a remark. The country reprocesses the spent fuel from its reactors (in France, but it is building domestic reprocessing facilities), has accumulated almost 10 tons of plutonium (and still has almost 160 tons in non reprocessed spent fuel), and has all the necessary know-how and technical skills

---

What could go wrong?, The National Interest, 3 August 2017, https://nationalinterest.org/blog/the-buzz/dead-russian-submarine-sitting-the-bottom-the-ocean-armed-21767 ).

[80] This military use of depleted uranium was evidently known and ready since a long time, but it is telling that it was adopted after the collapse of the Soviet Union.

[81] Xxx New disturbing data comes from Italy: 348 soldiers died from depleted uranium on Kosovo and Metohija, *Telegraph*, 14 December 2017, http://www.telegraf.rs/english/2919524-new-disturbing-data-comes-from-italy-348-soldiers-died-from-depleted-uranium-on-kosovo-and-metohija. Depleted uranium caused Italy soldiers' cancer - probe (4), *ANSA Latest News*, 7 February 2018, http://www.ansa.it/english/news/2018/02/07/depleted-uranium-caused-italy-soldiers-cancer-probe-4_0ceba3e0-f0da-484c-aba5-30522d0210d9.html.

[82] World Plutonium Inventories, *Bulletin of the Atomic Scientists*, Vol. 55, 1999, https://www.tandfonline.com/doi/abs/10.1080/00963402.1999.11460377; *Global fissile material report, nuclear weapon and fissile material stockpiles and production*, 2015, http://fissilematerials.org/library/ipfm15.pdf.

[83] J. Green, Can 'reactor grade' plutonium be used in nuclear weapons?, *Wise*, 6 June 2014, https://www.wiseinternational.org/nuclear-monitor/787/can-reactor-grade-plutonium-be-used-nuclear-weapons.
G. S. Jones, Reactor-Grade Plutonium and Nuclear Weapons: Exploding the Myths, *Nonproliferation Policy Education Center*, April 2018, http://npolicy.org/books/Reactor-Grade_Plutonium_and_Nuclear_Weapons/Greg%20Jones_Reactor-grade%20plutonium%20web.pdf.

[84] Victor Gilinsky, Marvin Miller, and Harmon Hubbard, A Fresh Examination of the Proliferation Dangers of http://npolicy.org/article.php?aid=172. In particular Appendix 3, Plutonium from Light Water Reactors as Nuclear Weapon Material points out that "The Department of Energy, in 1977, declassified the fact that an underground test had been conducted (in 1962) in which weapon grade Pu had been replaced with reactor grade Pu with successful results."

to build nuclear weapons in an extremely short time (it is named a *stand-by nuclear power*).

*Alert for "dirty bombs" (radiological dispersal device, RDD)* – Apart from military materials, a deep concern exists, and periodically re-emerges, that radioactive materials that can be found virtually in any country in the world could be used to build a radiological dispersal device (RDD), commonly known as a "dirty bomb". The concern is in particular the possible use of this material in a terrorist attack.

This recurrent concern had a recent case when a radioactive source was reported missing on past August 10 in Malaysia.[85] I take this case just as emblematic of a much more common problem. "According to reports from the Malaysian Atomic Energy Board, there have been more than 16 cases involving the theft or loss of radioactive material since the 1990s". The missing device is an industrial radiography unit with an iridium 192 isotope used for non-destructive testing. The incident caused concerns at the highest levels of the government, and it was discussed in the National Security Council of Malaysia. In the past, sophisticated dual-use technologies being manufactured being manufactured in the country had been a target and platform for nuclear smuggling. In the early 2000's, the A.Q. Kahn network used Malaysia as a starting point to produce and export materials to Lybia.

The problem are not only terrorists, as shows the case of a research institution in India which in 2010 improperly disposed of a gamma ray unit containing cobalt 60 seling it to scrap dealers who dismantled the equipment, and one of the scrap workers died, and six suffered radiation injuries.

It is important to mention that no international instrument binds countries to report the loss of control of radioactive sources and material to the IAEA or neighboring countries.

*Orphan radioactive sources* – Partly related with the previous problem is that of uncontrolled radioactive sources, called orphan sources. There have been several accidents over the past decades involving orphan radioactive sources or other radioactive material that were inadvertently collected as scrap metal that was destined for recycling. The melting of an orphan source with scrap metal or its rupturing when mixed with scrap metal has resulted in contaminated recycled metal and wastes.

## Uranium mining, and exploitation of poor populations

Finally, let us return to the front end of the nuclear cycle, the extraction of uranium, which was at the origin of the *Nuclear Age*. How was, and is, obtained the uranium?

Uranium is contained in ores, therefore it must be extracted and subsequently processed. Here there are other dramatic, and little-known, parts of the story. *Poor or exploited populations were always assigned to the mining of uranium ores, and they suffered dramatic health consequences*.

In the United States it was up to the Navajo population, whose region was rich in uranium ores, to make the major contribution to the implementation of the nuclear stockpile and industry. From 1944 to 1986, nearly 30 million tons of uranium ore were extracted from Navajo lands under leases with the Navajo Nation. Government and companies did not provide information on the risks, Navajo workers even suffered discrimination with respect to the white ones. The incidence of cancer and other related illnesses in the population is very high.[86] The problem of compensations has never been settled. Each

---

[85] F. Parada et al., Radioactive material is still missing in Malaysia: Cause for concern?, *Bulletin of the Atomic Scientists*, 14 September 2018.

[86] Shocking photos are easily available in internet. Some references are the following. D. Brugge et al., Exposure Pathways and Health Effects Associated with Chemical and Radiological Toxicity of Natural Uranium: A Review, *Reviews on Environmental Health*, Vol. 20, p. 177, 2005, https://www.degruyter.com/view/j/reveh.2005.20.3/reveh.2005.20.3.177/reveh.2005.20.3.177.xml. R. Billy, Navajo Nation faces ongoing risks from past uranium mining,

year the 16th of July the Navajo Nation celebrates the *Uranium Legacy Remembrance and Action Day*.

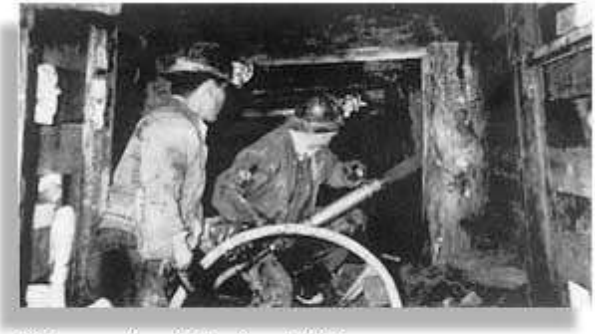
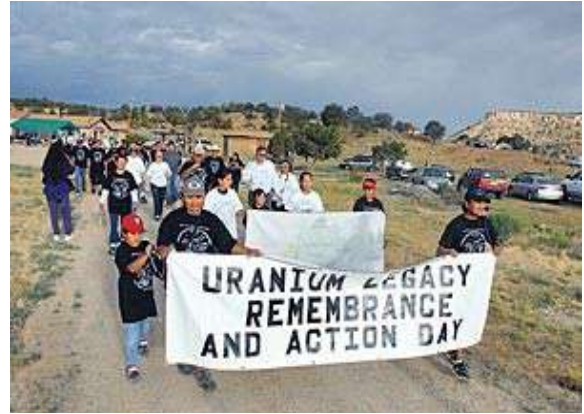

Navajo uranium miners, and Uranium Legacy Remembrance and Action Day.

France, after national uranium mines were exhausted, extracts it in Niger, where it exploits local population with similar consequences, and has moreover caused a wide and frightening radioactive pollution of the country.[87]

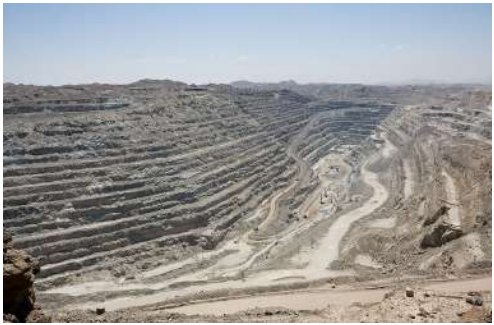
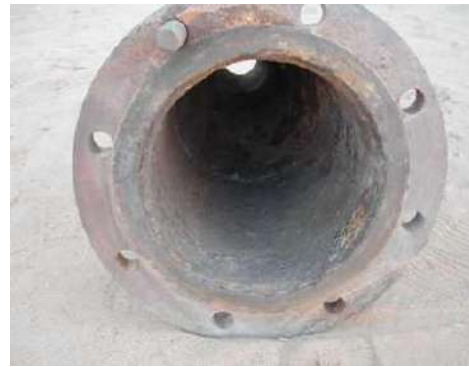

Niger: uranium extraction; people is accustomed to reuse materials, this pipe has a gamma activity 10 times higher than normal.

Moreover, France has another worst-kept, paradoxical (or maybe not) secret, since the extraction and processing of uranium from domestic mines left a widespread contamination throughout the country! The low activity waste spread in the country amounts to approximately 200-300 million tons. This scandal has been repeatedly denounced, for instance in an inquiry transmitted on 11th February 2009 by the channel France 3, with the title "Uranium, le scandale de la France contaminée"[88] (Uranium, the scandal of contaminated France).

To conclude, we are leaving to future generations an extremely heavy, costly, and dangerous burden, which could last for the foreseeable future of humankind. Or even longer!

---

https://swes.cals.arizona.edu/environmental_writing/stories/fall%202013/billy%20.html. B. Marcus, Toxic legacy of uranium mining in Native Southwest, *Liberation*, 23 June 2017, https://www.liberationnews.org/resource-extraction-of-the-american-indigenous-population-uranium/.

[87] Safety concerns dog French uranium mines in Niger, *The Guardian*, 15 October 2010, https://www.theguardian.com/world/2010/oct/15/niger-mining. N. Meynen, France's dirty little secret: nuclear pollution in Niger, *Meta*, 18 October 2017, https://metamag.org/2017/10/18/french-state-owned-company-creates-ecocide-in-niger-to-fuel-its-nuclear-plants/. A. Mohanti, Uranium in Niger: When a Blessing Becomes a Curse, *Geopolitical Monitor*, 19 April 2018, https://www.geopoliticalmonitor.com/uranium-in-niger-when-a-blessing-becomes-a-curse/.

[88] See to believe: https://www.youtube.com/watch?v=spum6lk6o4c. "Pendant quarante ans, les 210 mines d'uranium situées sur le territoire français ont alimenté l'industrie nucléaire et nos centrales. Depuis la fermeture des sites d'exploitation, de nombreuses populations vivent, sans le savoir, sur des terres contaminées par les déchets radioactifs.